\documentclass[fleqn,usenatbib]{mnras}

\usepackage{comment}

\usepackage{newtxtext,newtxmath}
\usepackage[ruled,vlined]{algorithm2e} 
\usepackage[T1]{fontenc}

\DeclareRobustCommand{\VAN}[3]{#2}
\let\VANthebibliography\thebibliography
\def\thebibliography{\DeclareRobustCommand{\VAN}[3]{##3}\VANthebibliography}

\usepackage{graphicx}	%
\usepackage{amsmath}	%
\usepackage{amssymb}
\usepackage{color,units}
\usepackage[dvipsnames]{xcolor}
\usepackage{float}
 \usepackage{capt-of}

\newcommand{\params}{\ensuremath{\theta}}
\newcommand{\likelihood}{\mathcal{L}}

\title[]{Massively parallel Bayesian inference for transient gravitational-wave astronomy}

\author[Rory J.E. Smith et al.]{
Rory J. E. Smith,$^{1,2}$\thanks{E-mail: rory.smith@ligo.org} Gregory Ashton,$^{1,2}$\thanks{E-mail: greg.ashton@monash.edu} Avi Vajpeyi $^{1,2}$\thanks{E-mail: avi.vajpeyi@monash.edu} and Colm Talbot$^{3,1,2}$\thanks{E-mail: colm.talbot@caltech.edu}
\\
$^{1}$School of Physics and Astronomy, Monash University, Vic 3800, Australia\\
$^{2}$OzGrav: The ARC Centre of Excellence for Gravitational Wave Discovery, Clayton VIC 3800, Australia\\
$^{3}$LIGO, California Institute of Technology, Pasadena, CA 91125, USA
}

\pubyear{2020}

\begin{document}
\label{firstpage}
\pagerange{\pageref{firstpage}--\pageref{lastpage}}
\maketitle

\begin{abstract}
Understanding the properties of transient gravitational waves and their sources is of broad interest in physics and astronomy. Bayesian inference is the standard framework for astrophysical measurement in transient gravitational-wave astronomy. Usually, stochastic sampling algorithms are used to estimate posterior probability distributions over the parameter spaces of models describing experimental data. The most physically accurate models typically come with a large computational overhead which can render data analysis extremely time consuming, or possibly even prohibitive.
In some cases highly specialized optimizations can mitigate these issues, though they can be difficult to implement, as well as to generalize to arbitrary models of the data. Here, we propose an accurate, flexible and scalable method for astrophysical inference: parallelized nested sampling. The reduction in the wall-time of inference scales almost linearly with the number of parallel processes running on a high-performance computing cluster. 
By utilizing a pool of several hundreds or thousands of CPUs in a high-performance cluster, the large wall times of many astrophysical inferences can be alleviated while simultaneously ensuring that any gravitational-wave signal model can be used ``out of the box'', i.e., without additional optimization or approximation. Our method will be useful to both the LIGO-Virgo-KAGRA collaborations and the wider scientific community performing astrophysical analyses on gravitational-waves. An implementation is available in the open source gravitational-wave inference library \texttt{pBilby} (parallel \texttt{bilby}).
\end{abstract}


\section{Introduction}
Gravitational-wave (GW) transients from merging binary black holes (BBHs) and binary neutron stars (BNSs) are now being routinely detected by the Advanced LIGO and Virgo detector network \citep{gracedb_public}. These compact-binary systems offer unprecedented means to study strong-field gravity \citep{LIGOScientific:2019fpa,PhysRevLett.123.011102, PhysRevLett.116.221101}, matter at supra-nuclear densities \citep{PhysRevLett.121.161101}, and stellar astrophysics \citep{LIGOScientific:2018jsj}. With data in the public domain \citep{Vallisneri:2014vxa}, methods to infer the properties of gravitational waves are of broad interest to various communities in physics and astronomy.

Bayesian inference is the standard framework for performing precision astrophysical measurements in transient GW astronomy \citep{LIGOScientific:2019fpa}. The output of Bayesian inference is two fold: $(i)$ posterior probability densities of astrophysical quantities encoded in gravitational waves, such as the masses and spins of binary black holes, and $(ii)$ an estimate of the probability of having observed the data under a particular hypothesis -- commonly called the \textit{evidence} -- which is used for hypothesis testing. Broadly speaking, there are three key ingredients to Bayesian Inference: $(i)$ experimental data, $(ii)$ a model, e.g., of the signal and noise components in interferometer strain data, and $(iii)$ an algorithm to efficiently explore the parameter space of the models, i.e., the astrophysical parameters of interest. Usually, stochastic sampling algorithms, such as nested sampling~\cite{skilling2006} or Markov Chain Monte Carlo (MCMC)~\citep{metropolis,hastings} are used due to the high dimensionality of the parameter spaces \citep{Veitch_2015, pycbcinf, Ashton:2018jfp}.

Generally, there are two classes of astrophysical inferences in transient GW astronomy. One class performs inferences on individual signals, e.g., from compact binary mergers. Here the source properties of individual transient signals are inferred, such as the masses, spins, source location etc.. Another class aims to infer the ensemble properties of particular GW sources. For example, the BNS merger rate, or the mass and spin spectrum of BBHs. This class of takes as input the inferences made on individual events, and is therefore known as \textit{hierarchical inference}, or population inference \cite{LIGOScientific:2018jsj}. Here, we focus on the former class of inferences, though the methods presented in this paper will also be applicable to population inferences.

Despite years of technical advances, Bayesian inference in GW astronomy remains challenging: high dimensional parameter spaces are difficult to explore efficiently, and the overall computational cost can be high. For example, when performing inferences on individual compact binary merger signals, the end-to-end analysis time -- commonly referred to as the \textit{wall time} -- can range between several hours to several weeks, months, or even years \citep{PhysRevD.94.044031}. 
The most expensive cases correspond to some of the most physically realistic and important analyses, but their high cost can be a major hurdle, or even roadblock, to astrophysical discovery. The large wall time of Bayesian inference on individual GW signals is the central issue that we address in this paper.
 
 In some cases, approximate methods can mitigate the large wall times. A class of approximations collectively known as reduced order methods \citep{PhysRevD.94.044031, Purrer:2014fza, Purrer:2015tud, Blackman:2017pcm, Varma:2018mmi, Canizares:2014fya} employ dimensionality reduction techniques to achieve low-latency parameter estimation. They generally find an approximation to signal models that is computationally more tractable than the underlying model. Algorithms such as RIFT \citep{Pankow:2015cra, Lange:2018pyp, Wysocki:2019} achieve rapid parameter estimation by pre-computing aspects of the inference problem in parallel and approximating expensive functions using cheaper interpolation methods.  Both these methods utilize an ``offline/online'' decomposition in which expensive computations are first performed offline -- possibly in parallel -- in order to facilitate fast online analyses using problem-specific approximations. Generally, these techniques all face the so-called ``curse of dimensionality'', i.e., they become exponentially more difficult to apply as the parameter space of models increases. Graphical processor units (GPUs) can, in some cases, reduce the cost of inference by more than an order of magnitude \citep{Talbot:2019okv}. However their utility is limited to a small class of GW signal models and so they cannot be used for general inference problems.

Our overall aim here is to provide a general framework for performing Bayesian inference in transient GW astronomy that significantly lowers the wall-time of data analysis. We are motivated by three considerations: $(i)$ accuracy, meaning the framework should produce statistically robust inferences; $(ii)$ flexibility, meaning it should be agnostic to the models and data being used; and $(iii)$ scalability, meaning it can handle a growing amount of work by adding computational resources. 

We show that wall-time of astrophysical inference on individual GW signals can be significantly reduced using a highly flexible, massively-parallel nested sampling algorithm deployed at scale on a high-performance CPU cluster. The reduction in wall-time scales almost linearly with the number of CPUs in the cluster. In some cases, our method reduces the wall-time from several years to around a week using the most physically complete GW signal models. Our method bridges the gap between a lack of available fast approximate methods, and the need to perform timely precision inference on individual GW events. The method meets the three criteria of accuracy, flexibiilty and scalability defined above. While our particular application is to inferences on individual GW signals, the method is agnostic to the particular inference problem and so will also be useful for reducing the wall time of hierarchical inferences. This represents a major advancement of techniques to mitigate the wall time of inference in gravitational-wave astronomy.  More broadly, the methods presented here should be useful to other fields in astronomy where the cost of inference is dominated by expensive calls to parameterized models of experimental data. 

The remainder of this paper is organized as follows. In Sec.~\ref{sec:BI} we give an overview of Bayesian Inference. In Sec.~\ref{sec:NS} we describe nested sampling and our parallelization scheme. In Sec.~\ref{sec:performance} we benchmark the performance of parallel nested sampling. In Sec.~\ref{sec:comparison} we discuss how parallel nested sampling compares to alternative methods for reducing the wall time of inference in transient GW astronomy. In Sec.~\ref{sec:extensions} we describe further applications of parallel nested sampling to GW astronomy. Finally, we give concluding remarks in Sec.~\ref{sec:conclusion}.

\section{Bayesian Inference}
\label{sec:BI}
Bayesian inference generally consists of two parts: \textit{Parameter estimation} and \textit{hypothesis testing} \citep{doi:10.1098/rstl.1763.0053, PhysRevLett.116.241102, PhysRevD.91.042003, Ashton:2018jfp, Pankow:2015cra}. Parameter estimation entails computing the posterior probability density of the source parameters given the experimental data, e.g., the masses and spins of binary black holes. Hypothesis testing entails computing the Bayesian ``evidence'': The probability of the data given an hypotheses. With the evidence, one can quantify the relative probability of the data under competing hypotheses, e.g., \textit{How much more probable is it that the data contain a signal with higher order mode content than a signal with only the leading-order quadrupolar mode?} \citep{collaboration2020gw190412} 

Inferences made about the astrophysics of individual gravitational-wave transients generally rely on $(i)$ a model for the underlying signal and possibly noise components of the data, and $(ii)$ a statistical description of noise processes in the data \citep{PhysRevD.91.042003,PhysRevLett.116.241102, Pankow:2015cra, Ashton:2018jfp}. Stochastic sampling algorithms -- such as nested sampling \cite{skilling2006}, or Markov Chain Monte Carlo \cite{metropolis,hastings} -- are employed to search the parameter space of the models and estimate posterior densities and evidences.

Bayesian inference relates the probability of model parameters $\theta$ to experimental data $d$, and an hypothesis for the data $\mathcal{H}$, via Bayes theorem: 
\begin{equation}
\label{eq:bayes}
    p(\theta|d, \mathcal{H}) = \frac{\pi(\theta| \mathcal{H})\,\mathcal{L}(d|\theta,\mathcal{H})}{\mathcal{Z}(d|\mathcal{H})}\,.
\end{equation}
Here, $p(\theta|d, \mathcal{H})$ is the \textit{posterior probability density} of the parameters $\theta$ given $d$ and $\mathcal{H}$; $\mathcal{L}(d|\theta,\mathcal{H})$ is the \textit{likelihood} of $d$ given $\theta$ and $\mathcal{H}$; $\pi(\theta| \mathcal{H})$ is the \textit{prior} probability of $\theta$; and $\mathcal{Z}(d|\mathcal{H})$ is the \textit{evidence} of $d$ given~$\mathcal{H}$. The posterior density is the target for \textit{parameter estimation}, while the evidence is the target for \textit{hypothesis testing}. 
Both the posterior and evidence can be estimated to high accuracy using nested sampling or thermodynamic integration \citep{PhysRevD.91.042003, pycbcinf, Ashton:2018jfp}. Assuming the priors can be defined, the primary input to inference algorithms is the likelihood function.

To motivate discussion of the computational cost of inference on modelled coalescing compact binary signals, we consider the usual likelihood function which describes the probability of interferometer data given $(i)$ an hypothesis that the data contain a signal plus Gaussian noise ($\mathcal{H}_S$), and $(ii)$ parameters $\theta$ which describe a model of the GW signal signal. This likelihood is the basis for most inferences on individual transient signals in GW astronomy \citep{PhysRevD.91.042003, Abbott:2018wiz, Ashton:2018jfp}, and constitutes the dominant cost of inference \citep{PhysRevD.94.044031, Pankow:2015cra}.   
The uncertainty on the data is due to the random noise component which is a stationary Gaussian process, coloured by the noise power spectral densities of the interferometers. The likelihood is:
\begin{equation}
\label{eq:cbc_likelihood}    
\mathcal{L}(d|\theta,\mathcal{H_S}) = \prod_{i}^{N_{\mathrm{det}}} \sum_{j}^{M} \frac{1}{2\pi S_{ij}}\exp\Bigg(-\frac{2}{T}\frac{|\tilde{d}_{ij}-\tilde{h}_{ij}(\theta)|^2}{S_{ij}}\Bigg)
\end{equation}
Here, $\tilde{d}$ and $\tilde{h}$ are respectively the Fourier transforms of the strain data and signal model, $S$ is the detector noise power spectral density and $T$ is the duration of the data in seconds. The product runs over the number of interferometers $N_{\mathrm{det}}$, and the sum runs over the number of frequency bins in Fourier transformed data/model $M$.

\subsection{Models and parameter spaces}

The choice of signal model $h(\theta)$ defines a \textit{particular} signal hypothesis. In practice many different signal models are often used to analyze a given signal \citep{LIGOScientific:2019fpa}. This can be to evaluate the significance of certain physics present in the signals, e.g., higher order modes, by computing the evidence of the data using models with and without higher order mode content \citep{collaboration2020gw190412}. Using multiple models can also serve to estimate systematic uncertainty in inferences that exists due to differences between models. Here we consider three fiducial signal models. For binary black hole analyses, we consider the models known as \texttt{IMRPhenomPv3HM}~\cite{Khan_2020} and \texttt{SEOBNRv4PHM}~\citep{SEOBNRv4PHM}. For binary neutron star analyses, we use an effective-precession model \texttt{IMRPhenomPv2NRT} \citep{Hannam:2013oca, Khan:2015jqa, Dietrich:2018uni, Husa:2015iqa, Dietrich:2017aum}. Due to the higher-order mode content, these BBH models are crucial for precision physics measurements on GWs from BBHs when the mass ratio of the systems is asymmetric, such as GW190412 \citep{collaboration2020gw190412}. Both models represent the current state-of-the art of binary black hole models that cover a large mass and spin range. They are also the most expensive BBH models.  \texttt{IMRPhenomPv2NRT} includes the effect of precessing spin on the heavier of the two bodies, and models the tidal deformability of neutron stars through two tidal deformability parameters. This model was used in the LIGO/Virgo analyses of GW170817 and GW190425, as well as in numerous other studies \citep{LIGOScientific:2019fpa}.

In addition to signal models, it is common to include models for the noise features. Typically, we model uncertainty of the data calibration \citep{Cahillane:2017jb}, and use a point estimate for the power spectral density (typically generated either using off-source data or an on-source estimation method \citep{Cornish:2014kda, Chatziioannou:2019dsz}).

The dimensionality of model parameter spaces can be highly variable. Astrophysical BBHs are described by 15 parameters (masses, spins, source location etc...), BNSs are described by an additional two parameters that describe the tidal deformability of the stars. The data-calibration model uses a set of amplitude and phases to model systematic uncertainty in the Fourier-domain data at a set of judiciously chosen frequency nodes. Typically, ten nodes are used per interferometer data set and the calibration model is described by 20 parameters (ten amplitudes and ten phases). Thus, data from a three-detector network are described by 75-77 parameters which must be inferred simultaneously.   

\subsection{The computational cost of inference}
There are two scales that determine the overall computational cost of Bayesian inference: $(i)$ The cost of evaluating parameterized models of the data, and $(ii)$ the rate of convergence of the sampling algorithms. We find it convenient to measure the computational cost in terms of CPU time, as this can be used to determine the wall time of the inference process. The cost of $(i)$ generally determines the CPU time of one iteration of a stochastic sampling algorithm, while $(ii)$ determines the overall CPU time required to complete the analysis.

The typical wall time can be estimated by first considering the total CPU time. To leading order, the CPU time $T_c$ of Bayesian inference scales like the average call-time of the data model $\langle T_m \rangle$, multiplied by the total number of calls to the likelihood function $N$ of the stochastic sampling algorithm
\begin{equation}
   T_c = N\langle T_m \rangle. 
\end{equation}
 We treat $N$ as an overall normalization which is typically $N\sim \mathcal{O}(10^7)$. When \textit{serial} sampling algorithms are used, the CPU time $T_c$ is equal to the wall time $T_w$. The average call-time $\langle T_m \rangle$ is strongly dependent on the complexity of the gravitational-wave signal models, and possibly models for the noise. 
 
 \subsubsection{Coalescing compact binary signal models}
 For a given model, $\langle T_m \rangle$ scales with the signal's bandwidth multiplied by its duration, which is equal to $M$ in the sum in Eq.~\ref{eq:cbc_likelihood} \citep{Canizares:2014fya}. The overall cost is set by the intrinsic complexity of the model \citep{PhysRevD.94.044031, Purrer:2014fza}. 
For signal models defined in the time domain, $\tilde{h}$ is computed by first evaluating the model in the time domain and subsequently taking the discrete Fourier transform.
 Time-domain signal models can be significantly more computationally expensive than those defined directly in the frequency domain. Many time-domain models require solving coupled ODEs to evaluate the signal at discrete times, see e.g., \cite{Pan:2013rra}. Together with the additional cost of the Fourier transform, the relative cost of using time domain models in inference can be between one to two orders of magnitude more expensive than frequency-domain models \citep{PhysRevD.94.044031, Purrer:2014fza}.

As a rule of thumb, more sophisticated models have higher $\langle T_m \rangle$. In practice, the range is broad: $\mathcal{O}(10^{-3}s) \lesssim \langle T_m \rangle \lesssim \mathcal{O}(1s)$. The lower limit corresponds to approximate frequency-domain signal models on short-duration binary black holes, e.g., \texttt{IMRPhenomPv2} \citep{Hannam:2013oca}. The upper limit corresponds to frequency-domain binary neutron star signal models, e.g., \texttt{IMRPhenomPv2NRT}, or complex time-domain binary black hole models that include spin precession effects and higher order modes, e.g., \texttt{SEOBNRv4PHM} and \texttt{IMRPhenomPv3HM}. Hence, for serial sampling algorithms, the wall time roughly ranges between $\mathcal{O}(1\,\mathrm{day}) \lesssim T_w \lesssim \mathcal{O}(1\, \mathrm{year})$. The upper limit presents serious hurdles, or possibly roadblocks, to using models with, e.g., higher order mode content, and two-body spin dynamics. 
 This problem will be compounded as GW detectors push their low-frequency sensitivity into the $5-10$Hz range because the in-band duration of observable signals will be up to an order of magnitude longer \citep{collaboration2019gravitational,Maggiore2020, reitze2019cosmic}.

\section{Parallel Nested Sampling}
\label{sec:NS}
Nested sampling is a stochastic-sampling method designed foremost to estimate the evidence $Z(d|\mathcal{H})$ \citep{skilling2006} in Eq.~(\ref{eq:bayes}), which is the primary ingredient in Bayesian hypothesis testing. As a byproduct, nested sampling also produces the posterior density $p(\theta|d,\mathcal{H})$. Importantly, nested sampling is scalable to high-dimensional and irregularly shaped parameter spaces \citep{Chopin_2010}. This affords a large degree of flexibility and ensures that nested sampling is well suited to extensions of the likelihood function in Eq.~(\ref{eq:cbc_likelihood}), e.g., by increasing the dimensionality of the parameter space to include parameters that model features of the noise, or parameters that describe signals in alternative theories of gravity.  

The evidence can be computed via the following integral:

\begin{eqnarray}
    Z(d|\mathcal{H}) &=& \int_{\Omega_{\theta}}\,d\theta\,\mathcal\pi(\theta)\,\mathcal{L}(d|\theta,\mathcal{H})\\
    &=& \int_{X=0}^{1}\,dX\,\mathcal{L}(d|X,\mathcal{H})\,\\
    &\approx& \sum_i \Delta X_i \likelihood(d|X_i,\mathcal{H})
    \label{eq:evidence}
\end{eqnarray}
The second line transforms the integral over the multi-dimensional parameter space $\theta$ into a one-dimensional integral over the \textit{prior mass} $dX=d\theta\,\pi(\theta)$. The quantity $\likelihood(d|X,\mathcal{H})$ is an iso-likelihood contour \citep{skilling2006}, i.e., it defines a boundary of constant likelihood within the prior volume X.

In practice, the inverse mapping $\theta(X)$ is not known, and so the integrals in Eq.~\ref{eq:evidence} cannot be performed analytically. Nested sampling estimates the evidence in  Eq.~\ref{eq:evidence} algorithmically.  Here we are agnostic to particular variants of nested sampling algorithms -- see, e.g., \cite{10.1093/mnras/stv1911, dynesty} -- because our aim is simply to remove one particular bottleneck that occurs due to the high cost of evaluating the models that enter the likelihood function. We therefore will describe parallel sampling in the context of one of the most basic variants of nested sampling known as ``static nested sampling'' \citep{dynesty}. 
We will not discuss the theory of nested sampling in depth, and we refer the reader to \cite{skilling2006, dynesty}.  However, we will find it useful to sketch the main algorithmic components of static nested sampling in order to introduce the parallelization scheme. There are three key components: $(i)$ prior sampling, $(ii)$ evidence estimation, and $(iii)$ obtaining posterior samples. The parallelization scheme enters into stage $(i)$. Before we describe the scheme, we briefly describe the three elements below. 

\subsubsection{Prior sampling}
Nested sampling estimates Eq.~\ref{eq:evidence} by drawing samples from the prior distribution. Samples are accepted subject to the constraint that those drawn on subsequent iterations have a higher likelihood than those on previous iterations. A key element is the set of \textit{live points}. The algorithm is seeded by drawing a number $K$ live points from the prior. These points are ranked from highest to lowest likelihood.  
The algorithm then proceeds by drawing samples $\theta_i$ from the prior on each iteration $i$ . The aim is to replace the live point with the lowest likelihood $\likelihood_{\min}$ on each iteration. Samples $\theta_i$ are accepted on each iteration subject to the constraint $\likelihood(d|\theta_i) \geq \likelihood_{\min}$. The sample associated with $\likelihood_{\min}$ is removed from the list of live points and added to a list of \textit{dead points}, and the new pair $\lbrace \theta_i, \likelihood(d|\theta_i)\rbrace$ is added to the list of live points. 
\subsubsection{Evidence estimation}
Once a sample has been accepted, the prior volume $X_i$ bounded by the likelihood $\likelihood(d|\theta_i)$ can be estimated as \citep{skilling2006} $X_i \approx [K/(K+1)]^i$, or equivalently $\ln X_i \approx -i/K$. With a set of likelihoods and an estimate for the change in prior volume $\Delta X_i = X_i - X_{i-1}$, the Riemann sum in Eq.~\ref{eq:evidence} can be computed on each iteration. The algorithm terminates when the change in the (log) evidence is below some user-defined threshold: $\Delta \ln Z = \ln Z_i - \ln Z_{i-1} \leq \epsilon$. \

\subsubsection{Posterior samples}

Once the algorithm has terminated, the posterior can be estimated as follows. Because the evidence is the integral of the \textit{un-normalized posterior} density, we must have 
\begin{equation}
Z \approx \sum_i \Delta X_i\likelihood(d|\theta_i) = \sum_i p(\theta_i) \,, 
\end{equation}

where $p(\theta_i)$ is an ``importance weight'' which represents an estimate of the un-normalized posterior density at sample point $\theta_i$: $p(\theta_i) \approx \likelihood(d|\theta_i)\pi(\theta)\Delta\theta_i$. The importance weights can then be used to approximate the posterior

\begin{eqnarray}
    p(\theta|d,\mathcal{H}) &\approx&  \frac{\sum_i p(\theta_i)\delta(\theta-\theta_i)}{\sum_i p(\theta_i)}\,\\
    &=& Z^{-1}\sum_i p(\theta_i)\delta(\theta-\theta_i)
\end{eqnarray}

\subsubsection{Parallel prior sampling}

In practice, a bottleneck arises when drawing prior samples to update the live points. This is because drawing samples requires evaluating the likelihood constraint, and hence the likelihood function, which is computationally expensive.  This bottleneck can be alleviated by parallelizing the prior-sampling step.

The parallel variant of static nested sampling is shown in Alg.~\ref{alg:p-nest}. The only difference to \textit{serial} static nested sampling is that samples will be drawn from the prior in parallel on each iteration. This is possible because each iteration of the nested sampling algorithm is independent of the state of the algorithm on previous iterations, i.e., a series of draws from the prior is equivalent to the same draws being made simultaneously.    Intuitively, if we were able to achieve perfect scaling, we would be able to advance the state of the nested sampling algorithm by exactly a factor of $n$ on each iteration because we could make $n$ live-point updates simultaneously. The parallel sampling procedure is straightforward to implement on $n_{\text{cores}}$ CPU cores via Message Passing Interface (MPI) \citep{DALCIN20111124}. We use a head/worker model\footnote{Note that this is usually called a ``master/slave'' model} where the ``head'' node organizes live/dead points, and estimates the evidence, while the $n_{\text{cores}}-1$ ``worker'' nodes find new live points. On each iteration $i$, a CPU $j$ evolves the same lowest-likelihood live point $\likelihood_{\min}$. A sample $\theta_{i,j}$ is drawn from the prior and is accepted subject to the usual constraint $\likelihood(d|\theta_{i,j}) \geq \likelihood_{\min}$, or rejected otherwise. Once the $n^{\prime}\leq n_{\text{cores}}-1$ samples have been gathered, they can be used to update the list of live and dead points. 
We can iteratively replace $\lbrace \likelihood_{\min}, \theta(\likelihood_{\min})\rbrace$ with $\lbrace \likelihood(d|\theta_{i,j}), \theta_{i,j}\rbrace_{j=1}^{n^{\prime}}$. 
In principle, we could let each of the workers continue sampling until they all find a valid sample point, however, this would create a sampling bottleneck whereby the list of live points cannot be updated until the least efficient worker returns a sample.

The probability of drawing a point between two iso-likelihood contours scales like the inverse of the volume contained between the contours. As such, parallel prior sampling will not in general draw samples that are guaranteed to be accepted as new live points, and hence we cannot expect to achieve linear scaling with the number of CPUs, $n_{\text{cores}}$.  We quantify the overall improvement in efficiency by a speedup factor which is a function of the number of live points and cores. The scaling relation for the speedup $S$ is \citep{10.1093/mnras/stv1911} 

\begin{equation}
\label{eq:scaling}
    S(n_{\text{cores}}, n_{\text{live}}) = n_{\text{live}}\ln(1+n_{\text{cores}}/n_{\text{live}}).
\end{equation}

The expected wall time of inference using parallel nested sampling therefore scales as

\begin{eqnarray}
    T_w(n_{\text{cores}}, n_{\text{live}}) &=& \frac{N}{S(n_{\text{cores}}, n_{\text{live}})}\langle T_m \rangle\\
    &=& \frac{T_c}{S(n_{\text{cores}}, n_{\text{live}})}
\end{eqnarray}

\begin{algorithm*}
\SetKwBlock{DoParallel}{do in parallel}{end}
    \SetAlgoLined
    \tcp{Initialize a pool of n CPUs}
    \tcp{Initialize live points}
    \DoParallel{Draw $K$ ``live'' points
    $\lbrace \theta_1, \dots, \theta_K \rbrace$ 
    from the prior $\pi(\theta)$ \\}
    \tcp{Main sampling loop}
    \While{${\rm stopping\:criterion\:not\:met}$}{
        Compute the minimum likelihood $\likelihood^{\min}$ among the
        current set of live points \\
        \tcp{Parallel sampling step on "worker nodes"}
        
        \DoParallel{
        Draw $n-1$ samples $\lbrace\theta_i\rbrace_{i=1}^{n-1}$ from the prior\\
        Accept $n^{\prime}$ samples subject to the constraint
        $\likelihood(\theta_i) \geq \likelihood^{\min}$, 
        otherwise discard \\
        }
        
        \tcp{Gather parallel samples on "head node"}
        
         \For{$i=1\, \mathrm{to\,} n^{\prime}$}{
        Add the $k^{th}$ live point $\params_k$ associated with $\likelihood^{\min}$
        to a list of ``dead'' points \\
        Replace $\theta_k$ with $\theta'_i$ \\
        Compute the minimum likelihood $\likelihood^{\min}$ among the
        current set of live points \\
        
        }

        \tcp{Check whether to stop}
        Evaluate stopping criterion \\
         \tcp{Check whether to update prior sampling method/parameters}
    Evaluate bounding distribution\\
    }
   
    \tcp{Add final live points}
    \While{$K > 0$}{
        Compute the minimum likelihood $\likelihood^{\min}$ among the
        current set of live points \\
        Add the $k$th live point $\params_k$ associated with $\likelihood^{\min}$
        to a list of ``dead'' points \\
        Remove $\params_k$ from the set of live points \\
        Set $K = K - 1$ \\
    }
    \caption{Static Parallel Nested Sampling}
    \label{alg:p-nest}
\end{algorithm*}

\section{Performance tests and results}
\label{sec:performance}
Parallel nested sampling is theoretically capable of reducing the wall-time according to the scaling relation in Eq.~\ref{eq:scaling}. We first compare the empirical scaling to the theoretical expectation. Secondly, we determine that our implementation of parallel nested sampling yields unbiased estimates of posterior densities.

In order to determine the speedup scaling, we measure the wall time of the binary black hole merger event GW150914 as a function of the number of CPU cores, keeping the number of live points fixed. This benchmark test should provide a generic scaling relation for the reduction in wall time, provided the likelihood function dominates the overall cost of inference and other costs are negligible. As such, it should be applicable to determine the speedup and reduction in wall time of analyses on other GW events and, e.g., hierarchical inference studies.

To demonstrate that the method produces unbiased posteriors, we perform a binary black hole ``injection campaign'': we create 100 synthetic BBH merger signals which we add into Gaussian, stationary noise colored with the aLIGO and Virgo PSDs. We then test the quality of the inferred posterior probability densities using a PP test.

\subsection{Scaling relation}

\subsubsection{Implementation}
We use the gravitational-wave inference package \textit{parallel bilby} (\texttt{pBilby}) to analyze the gravitational-wave event GW150914. Parallel nested sampling (Alg.~\ref{alg:p-nest}) is implemented in \texttt{pBilby} via the \texttt{dynesty} nested sampling library. Communication between nodes is accomplished using MPI through the python package \texttt{mpi4py} and \texttt{schwimmbad}. We analyze 4s of strain data containing the GW150914 from the LIGO-Hanford and LIGO-Livingston observatories. We use a minimum and maximum frequency of $20$Hz and $1024$Hz respectively. The data, noise PSD, calibration model and prior ranges were taken from the Gravitational-wave Open Science Center \citep{collaboration2019open}.

For the gravitational-wave likelihood function, we use two models, one for the BBH merger signal, and another for the calibration of the data. We use the gravitational-wave signal model \texttt{IMRPhenomPv3HM} and a data calibration model from \cite{Cahillane:2017jb}. The signal model is a cutting edge binary black hole model which includes the effects of spin-precession, and higher order gravitational-wave modes. The computational cost is typical of the current generation of signal models. The data calibration model is the standard model used in LIGO/Virgo analyses on compact binary mergers. Thus, the wall-time measurements will be indicative of the actual run times for real LIGO-Virgo-KAGRA analyses. In total, the model parameter space is 55-dimensional: 15 astrophysical parameters describe the gravitational-wave signal, and 40 describe the data calibration.

Our analyses use 2000 live points, which we have found is robust for inferences on BBH signals including a data-calibration model. To effectively bound the prior distribution and improve convergence, we use multi-elipsoid bounding distributions \citep{Feroz:2008xx} implemented in \texttt{dynesty}.  
To ensure we generate prior samples efficiently, we use a modified version of the MCMC proposal distribution implemented in \texttt{dynesty}. We fix the number of random walks in the MCMC to ensure that workers are synchronized and the cores are properly load balanced.   

To measure the scaling we record the wall-time of running \texttt{dynesty} with parallel prior-sampling on GW150914 using $n_{\text{cores}} = (16, 64, 320, 640)$ CPU cores to draw prior samples in parallel on each iteration. Note, that because one CPU process is reserved as the ``head'' process while the others draw samples, the number of CPUs drawing samples in parallel is $n_{\text{cores}}-1$. We perform five independent runs for each $n_{\text{cores}}$ to get a measure of the typical variation in wall times. From the wall-times, we can directly compute the scaling as a function of $n_{\text{cores}}$ which we can compare to Eq.~\ref{eq:scaling}. All of the runs were performed on Intel Xeon E5-2660 (Sandybridge) CPUs with a 2.2GHz clock rate. Nodes are networked via non-blocking QDR infiniband.

\subsubsection{Results}
\begin{center}
\begin{figure}
\includegraphics[width=\linewidth]{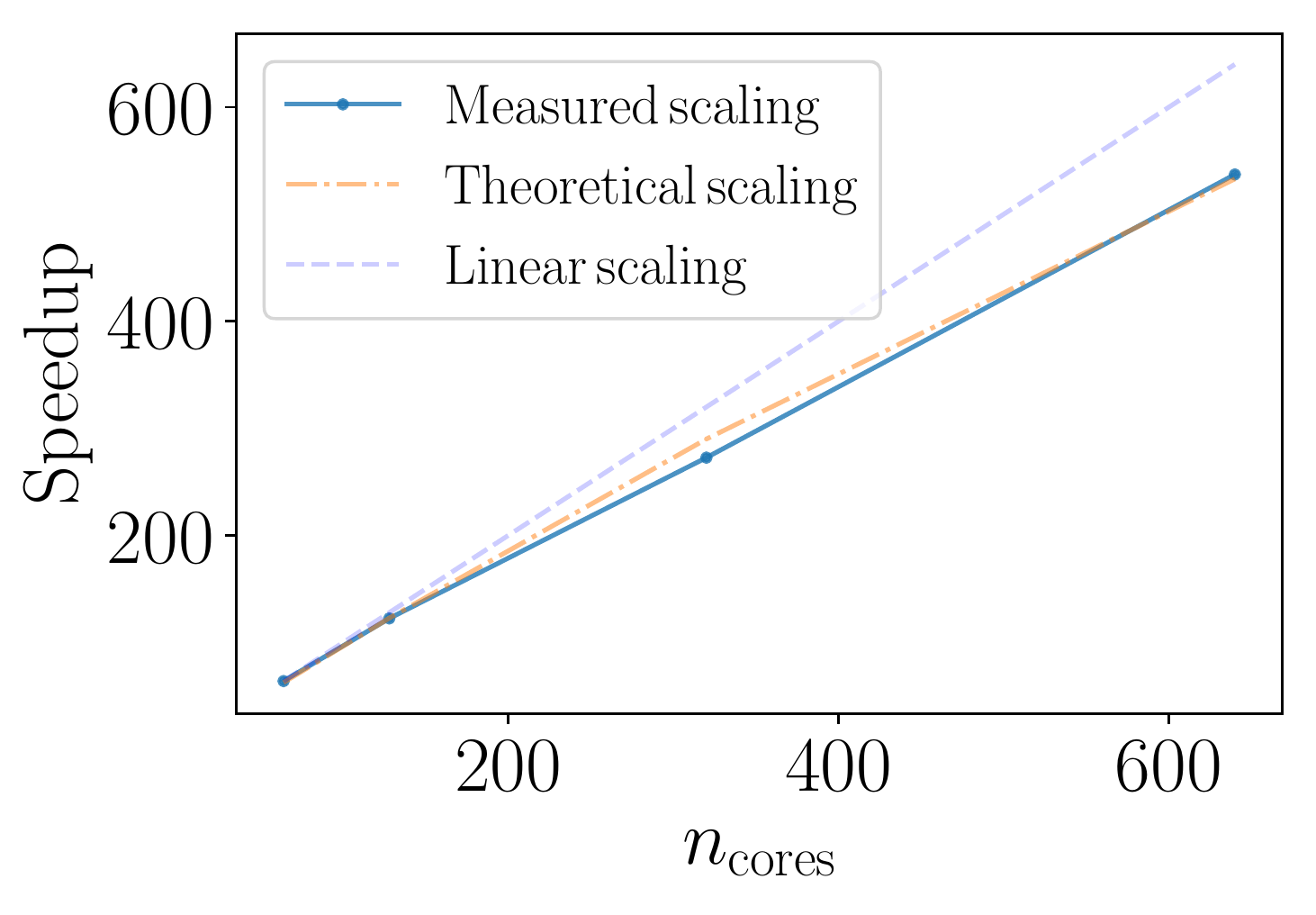}
\caption{Speed up factor Eq.~(\ref{eq:scaling}) vs number of CPUs for a fixed number of live points ($n_{\mathrm{live}}=2000$).}
\label{fig:scaling}
\end{figure}
\end{center}
The measured speedup-scaling is shown in Fig.~\ref{fig:scaling}.
We find excellent agreement with the theoretical prediction for the scaling for a fixed number of live points and variable number of CPU cores. The theoretical scaling curve is computed using Eq.~\ref{eq:scaling}. Because the sampling time is dominated by the cost $\langle T_m \rangle$ of evaluating the models that enter into the likelihood function, the scaling can be thought of as an effective decrease in the average wall-time of evaluating these models. Provided one is in the regime where the dominant cost of inference is $\langle T_m \rangle$, we expect that the scaling will hold when the number of CPU cores and/or the number of live points is increased. 

Moreover, as we have argued, the scaling does not depend on the \textit{particular} choice of models, likelihood functions, or even the type of data being analyzed. Thus, the scaling relation should be broadly applicable to a range of inference problems in astronomy where the cost of parameterized models dominates the cost of inference, e.g., population/hierarchical inference studies in GW astronomy. We also note that the scaling may be independent of the actual variant of nested sampling algorithm, provided the are compatible with the same or similar parallel prior-sampling methods.

\setlength{\tabcolsep}{1ex}
\begin{table*}\centering
\begin{tabular}{@{}rrrrrrrrrr@{}}
& \multicolumn{3}{c}{IMRPhenomPv3HM} & \multicolumn{3}{c}{SEOBNRv4PHM} & \multicolumn{3}{c}{IMRPhenomPv2NRT}\\\hline
Number of CPUs&$16$&$64$&$640$&$16$&$64$&$640$&$16$&$64$&$640$\\ \hline
GW150914 & 3.9 d & 23.3 hr & 2.8 hr & \textit{83.7} d & \textit{21.2} d &2.5 d&--&--&--\\
GW190425 & -- & -- & -- & --- & ---& ---& \textit{30.7} d& \textit{7.8 d} & 22 hr \\
GW190412 & \textit{60.3 d} & \textit{15.3 d} &  1.8 d& \textit{2.9} yr & \textit{276.1} d& 11.53 d& ---& --- & ---
\end{tabular}
\caption{Wall times for selected events using $n_{\text{cores}}=(16, 64, 640)$ CPUs. Measured wall times are non-italicized and estimated wall times are \textit{italicized}.}
\label{table:examples}
\end{table*}
 
\subsection{Example wall-time reduction} 
In Table \ref{table:examples} we show representative wall times for running parallel nested sampling on the gravitational-wave events GW150914, GW190425 and GW190412. These events correspond to a short-duration binary black hole merger, a binary neutron star merger, and a long-duration ($\sim$ 10s) binary black hole merger. Non-italicized wall times in Table~\ref{table:examples} are measured and italicized wall times are estimated from the measured values.  We consider the waveform families \texttt{IMRPhenomPv3HM, SEOBNRV4HM and IMRPhenomPv2NRT}. For each event analysis, we determine the wall time using $n_{\text{cores}} = (16, 64, 640)$. We use the same analysis set up as in Sec.~\ref{sec:performance}. Prior ranges are taken from \cite{collaboration2019open}. 

For analyses on GW150914-like systems, we determine that run times can be reduced to around 2.8 hours when using 640 CPU cores, down from 3.9 days using 16 cores when using \texttt{IMRPhenomPv3HM}. The scaling shown in Fig.~\ref{fig:scaling} is based on these measurements. Analyses on GW190412 are expected to scale similarly as the likelihood function will be an even more dominant cost due to the increased duration of the data. We find that for \texttt{IMRPhenomPv3HM} the analysis time can be reduced to 1.8 days on 640 cores, down from 60.3 days on 16 cores. For \texttt{SEOBNRv4PHM}, the analysis time can be reduced to 11.5 days on 640 cores, vs 2.9 years on 16 cores. These cases are particularly relevant to LIGO-Virgo data analysis as reduced order methods are not yet available for these signal models. For binary neutron star analyses on GW190425-like systems, we show that wall times can be reduced to 22 hrs on 640 cores vs 30.7 days on 16 cores when using \texttt{IMRPhenomPv2NRT}.

\subsection{Sampling accuracy}
Our goal here is to produce a metric which measures the accuracy of the estimates of posterior density produced by the algorithm. Because the output of an analysis is a set of PDFs, any bias in a single set is hard to gauge. 
We therefore test the quality of an ensemble of posterior PDFs. We quantify sampling accuracy using a P-P test. We generate artificial data sets containing GW signals and LIGO-like noise. 
Our expectation is that the true parameter values should fall within the X$\%$ credible region $X\%$ of the time, signifying that our posterior densities are unbiased. For all estimated parameters, the P-P test computes the fraction of events for which the injected signal parameters fall within the $X\%$ credible interval, and assigns a p-value to the outcome. 

We use the likelihood function in Eq.~(\ref{eq:cbc_likelihood}) which contains 15 free parameters which describe the BBH signals. We do not consider a model of the data calibration. We use the same run configuration described in Sec.~\ref{sec:performance}. Signals are injected with parameters drawn from the prior used for the analysis of GW150914. We analyze 100 synthetic gravitational-wave signals.

\begin{center}
\begin{figure}
\includegraphics[width=\linewidth]{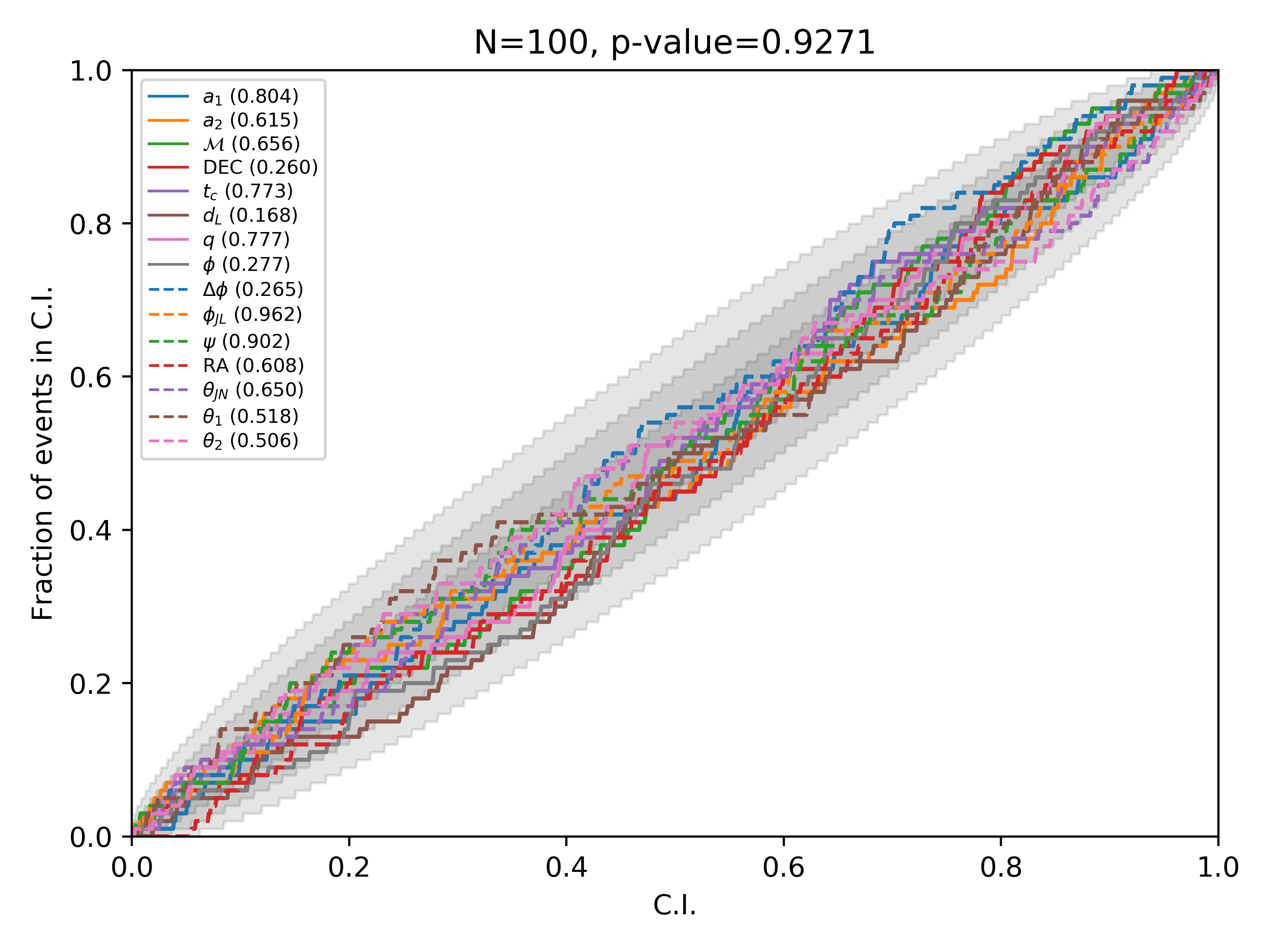}
\caption{Credible interval (CI) vs the fraction of events in a particular CI for the 15 parameters that describe 100 simulated BBH merger signals. Perfect scaling results in all curves falling along the horizontal. The grey region shows the 1-2-3$\sigma$ uncertainty regions for the distribution of curves.}
    \label{fig:accuracy}
    \label{fig:PP}
    \end{figure}
\end{center}

The P-P plot is shown in Fig.~\ref{fig:PP}. The $x-$ and $y-$axes are respectively the credible interval (CI) and the fraction of events in a particular CI. Perfect scaling results in all curves falling along the horizontal. The grey region shows the 1-2-3$\sigma$ uncertainty regions for the distribution of curves. We find that our results are consistent with perfect scaling at the p-value $p=92.7\%$ level, implying that our posterior density estimates are robust. Thus, parallel nested sampling is accurate.

\section{Comparison to alternative methods}
\label{sec:comparison}
A class of techniques known collectively as ``reduced order methods'' have been successful at reducing the cost of inference using particular signal models by up to a factor of around 300, and are employed by the LIGO Scientific Collaboration in production-level analyses \citep{PhysRevD.94.044031, Purrer:2014fza, Purrer:2015tud}. One class known as a reduced order model, utilizes a problem-specific down-sampling of waveforms at judiciously chosen time or frequency nodes. The full waveform can be reconstructed together with an efficient global interpolation method to up-sample the waveforms to the full time or frequency space. A further class of methods known as reduced order quadratures exploit the reduced order model repesentation to compress the number of terms in the likelihood function Eq.~\ref{eq:cbc_likelihood}, effectively performing inference on compressed data. However, they are difficult to apply broadly to all classes of signal models, in particular to fully-precessing time-domain signal models, due to the curse of dimensionality, see, e.g., \cite{PhysRevX.4.031006,Blackman:2017pcm, Varma:2018mmi}. Moreover, they typically require highly specialized knowledge to construct and as such may not be readily utilized by the larger physics and astronomy communities. Nevertheless, given the current trajectory of research on reduced order methods, it seems likely that in the future they will exist for fully-precessing time-domain models. While we have not explored the possibility of combining parallel nested sampling with reduced order model waveforms or reduced order quadratures, it seems probable that they could further reduce the wall time of inference by between one and several orders of magnitude. Combining reduced order methods with parallel nested sampling therefore merits further attention. 

A method known as ``relative binning'' \cite{zackay2018relative} also exploits a reduced-order decomposition of the likelihood function, and has been shown to reduce the cost of likelihood/mode evaluations by a factor of around $10^4$ for binary neutron star mergers -- roughly an order of magnitude greater than the most efficient reduced order models. Relative binning utilizes ``summary data'' which captures sufficient information about how gravitational waveforms smoothly change over parameter space with respect to a fiducial waveform. This data is around an order of magnitude less than the level of down sampling achieved by reduced order models. The study in \cite{zackay2018relative} was limited to frequency-domain waveforms, though in principle it can also be applied to time-domain waveforms making the method fairly flexible. One drawback is that it requires that the summary data -- the complex-valued gravitational-wave strain at well-chosen frequency or time bins -- can be directly accessed. While this data can be easily accessed for frequency or time domain waveforms that admit closed-form expressions, e.g., \cite{IMRPhenomP}, it cannot be accessed for waveforms that do not. For example, in order to access the time-domain strain at a set of sparsely separated time bins for models such as \texttt{SEOBNRv4PHM} would still require solving the waveform at all intermediate time bins. This is because many time-domain waveform models require evolving the orbital dynamics via a set of coupled ODEs. Nevertheless, assuming that this issue can be overcome, e.g., with reduced order models, relative binning may be able to offer genuinely low-latency inference with or without parallel nested sampling. As with reduced order methods, we believe that research combining relative binning with parallel sampling is warranted.

Other methods offer various degrees of parallelism: Monte Carlo methods \citep{PhysRevD.91.042003} have been employed to facilitate using expensive models such as SEOBNRv3 \citep{LIGOScientific:2019fpa}. However, the efficiency of the algorithm is typically poor $\textit{and}$ the model is expensive. These two issues compound in such a way as to make the wall time \textit{and} CPU time are large\footnote{For example, the analyses on the GWTC-1 event GW170608 used 120 parallel MCMC chains running continuously for around two months \cite{eve_comm}}. 
Parallel ``grid-based'' methods such as RIFT \citep{Lange:2018pyp, Pankow:2015cra, Wysocki:2019} precompute certain aspects of the model/likelihood space in parallel in combination with interpolation techniques to estimate the posterior density and evidence. Importantly, RIFT can be used to estimate posteriors and evidences using, e.g., a relatively sparse set of numerical relativitiy simulations. This is achieved by evaluating the likelihood at specific points in parameter space which can then be interpolated across the domain of high posterior support. While this method offers significant advantages over pure sampling-based methods, the complexity of grid-based interpolation methods scales unfavourably with the dimensionality of the parameter space, i.e., it suffers from the curse of dimensionality. For this reason, it will be difficult to scale grid-based methods to the levels required by general problems in inference in GW astronomy.

An embarassingly parallel method known as likelihood reweighting \citep{Payne:2019wmy,kish}
produces posterior samples and evidences from a target likelihood function, i.e., the one of interest, by leveraging samples and evidences obtained using a reference likelihood function that is computationally cheaper to evaluate \citep{2018arXiv180904129E}. Reweighting is efficient when the target and reference posteriors are similar. For instance, analyses using higher-order mode models on the GWTC-1 events can generate posterior samples with between $7\% - 60\%$ efficiency \citep{Payne:2019wmy}. However, there are two drawbacks. First, when the target and reference likelihoods diverge, the overall efficiency can be poor. For loud events, e.g., SNR$\sim 50$, the efficiency can be around $0.1\%$ \citep{Payne:2019wmy} meaning that many thousands of reference analyses have to be performed to generate a satisfactory number of effective samples through reweighting. In practice this can lead to a very high overall CPU time, though due to the embarassingly parallel nature of the problem, a low wall time. Secondly, the choice of a good reference likelihood is not always obvious. Here, a trade off between accuracy and speed has to be made, and several of the fastest waveform models have restrictions in, e.g., mass ratio and spin \citep{PhysRevD.94.044031}, which could make the target and reference likelihoods diverge in regions of parameter space, thus introducing a source of inefficiency in the reweighting procedure. 

Lastly, we consider the use of GPUs. GPUs can accelerate aspects of the inference problem that are embarassingly parallel. In particular, many frequency-domain GW models admit closed form expressions and hence the model at each frequency bin can be evaluated in parallel. In \cite{Talbot:2019okv} the authors demonstrate that the cost of evaluating frequency-domain waveform models can be accelerated by a factor of $\sim 50$ using a single GPU.
We note that the method described here can be used to distribute sampling over a pool of GPUs to obtain further acceleration.
A clear drawback of GPU acceleration is that it is unlikely to be able to accelerate models which are computed by first solving couple ODEs, e.g., most time-domain models. These models must be evolved iteratively, and so the full time series cannot be evaluated in parallel in contrast to their frequency-domain counterparts. 

\section{Further applications}
\label{sec:extensions}
We have focused on individual event inference problems where the only free parameters are those of signals described by (approximations to) General Relativity, and data calibration. Nested sampling methods have been shown to be robust for estimating evidences and posteriors in parameter spaces that have many tens to hundreds of parameters \citep{Allison:2013npa, Feroz:2008xx}. Thus, our results demonstrate that provided the inference problem is dominated by the cost of the likelihood function, then parallelized nested sampling will offer comparable speedups for inferences in which the models and parameter spaces are significantly larger and more complex than those which we have considered. For example, it is increasingly common for analyses to estimate not just signal parameters but also those of models for the noise power spectral density \citep{Cornish:2014kda}. Additionally, signal models in alternative theories of gravity are parameterized by many more than the 15-17 parameters which describe binary-merger signals in general relativity \citep{TheLIGOScientific:2016src}. Thus, our method is extendable to a wide class of important (astro)physical analyses on individual GW events. 

In addition, to inference on individual GW events, parallel nested sampling will be useful in population (hierarchical) inference studies which estimate ensemble properties of GW events, such as the mass spectrum of binary black holes. In population inference, posterior samples from many events are combined self-consistently to infer information about the underlying distribution from which the samples were drawn. Typically, the cost of population inference scales like the number of samples per event multiplied by the number of events \citep{Talbot:2019okv}. In these problems, the cost of evaluating parameterized models dominates the cost of inference, as in inference on individual GW events considered in this paper.  This implies that the cost of population inference studies should be reduced according to the scaling in Eq.~\ref{eq:scaling}. As such, our method may be important when the number of events becomes very large, e.g., as LIGO/Virgo/KAGRA observe many hundreds of events. We note that GPU acceleration has already been shown to accelerate population inference by between one and two orders of magnitude \citep{Talbot:2019okv}. As such, parallel nested sampling may not be necessary until the volume of data required for population inference exceeds the memory capacity of GPUs.

Parallelized nested sampling may also serve as a useful tool in tackling inference on signals as seen by third-generation detectors, e.g., Einstein Telescope and Cosmic Explorer \cite{Maggiore2020,reitze2019cosmic}. Astrophysical analysis for these instruments will be significantly more complex:  many signals will be in-band simultaneously, and signals may be in band for up to several tens of minutes \cite{Sathyaprakash:2012jk, Reitze:2019iox, Evans:2016mbw}. Thus methods which can alleviate aspects of the wall-time of inference will be valuable as the demands and complexity of data analysis increase. As we note in Sec.~\ref{sec:comparison}, a combination of parallel nested sampling together with reduced order, or relative binning techniques, could lead to dramatic performance improvements. These could be crucial to inference with third-generation detectors, or space-based detectors such as LISA \cite{Amaro_Seoane_2012}, where in-band signals may be observable for several hours, to many days or weeks.

Beyond transient GW astronomy, parallel nested sampling and our \texttt{pBilby} code should have  applications throughout astronomy and astrophysics. For example, the serial variant of \texttt{pBilby} has been used in a number of studies in pulsar and radio astronomy \cite{Lower_2020, Jiang_2020, zhu2020unambiguous, Cho_2020}. The methods presented in this paper should offer greater scalability of data analysis in these fields.

\section{Conclusion}
\label{sec:conclusion}
Parallelized nested sampling, deployed at scale on a high-performance CPU cluster, reduces the wall-time of inference according to Eq~(\ref{eq:scaling}). It does not approximate either gravitational-wave signal models, or the statistical properties of the data; is accurate, flexible, scalable, and easy to implement. As such, is can be used in a broad variety of inference analyses. We have demonstrated reductions in wall time from several years to several days for realistic LIGO-Virgo analyses that use cutting-edge gravitational-wave signal and data-calibration models. 

We have argued that the measured speedup achieved by parallel nested sampling should apply irrespective of the type of data or models being used, provided the dominant cost of inference stems from expensive calls to likelihood/model functions. As such, our method -- and code, \texttt{pBilby} -- should be useful for other expensive inference problems, such as hierarchical inference in gravitational-wave astronomy.

While potentially computationally expensive, parallel nested sampling nonetheless affords greatly expedited inferences on gravitational waves provided one has access to a high-performance computer cluster. Given the increasing availability of clusters, together with cloud-computing resources, parallelized nested sampling should be a useful tool to both the LIGO-Virgo-KAGRA Collaboration, as well as to independent research groups in astronomy more broadly. 

\section{Acknowledgements}
This work is supported through Australian Research
Council (ARC) Centre of Excellence CE170100004. The analyses presented in
this paper were performed using the supercomputer
cluster at the Swinburne University of Technology (SSTAR). This document has LIGO Document number P1900255-v1. We would like to thank Mathew Pitkin, Roberto Cotesta, Simon Stevenson, Serguei Ossokine and Scott Coughlin for extensive testing of \texttt{pBilby}, and Eve Chase for providing information about the GWTC-1 analyses performed using SEOBNRv3. Thanks also to the SSTAR system admins for their support with all things MPI and for their patience. We are grateful for insightful comments from Vivien Raymond, Eve Chase, Richard O'Shaughnessy, Moritz Hubner, Michele Vallisneri, Alessandra Buonanno, Vicky Kalogera and the LIGO-Virgo Parameter Estimation and Coalescing Compact Binary working groups.  Additional thanks to Joshua Speagle for pointing out the scaling relation for parallel nested sampling. This research has made use of data, software and/or web tools obtained from the Gravitational Wave Open Science Center (https://www.gw-openscience.org), a service of LIGO Laboratory, the LIGO Scientific Collaboration and the Virgo Collaboration. LIGO is funded by the U.S. National Science Foundation. Virgo is funded by the French Centre National de Recherche Scientifique (CNRS), the Italian Istituto Nazionale della Fisica Nucleare (INFN) and the Dutch Nikhef, with contributions by Polish and Hungarian institutes.


\bibliographystyle{mnras}
\bibliography{biblio}

\bsp	
\label{lastpage}
\end{document}